\documentclass[preprint,12pt]{elsart}
\usepackage{graphicx,cite,url}
\usepackage{amssymb}
\usepackage{amsmath}
\usepackage{upgreek}
\usepackage{color}
\journal{Physics Letters B}

\newcommand{\affuni}[2]{Dipartimento di Fisica dell'Universit\`a #1, #2, Italy.}
\newcommand{\affinfn}[2]{INFN Sezione di #1, #2, Italy.}

\def\ifm#1{\relax\ifmmode#1\else$#1$\fi}  \def\to{\ifm{\rightarrow}}  
     \def\x{\ifm{\times}}      
\def\up#1;{$^{#1}$}  \def\dn#1;{$_{#1}$}      
    
\def\pt#1;#2;{\ifm{#1\x10^{#2}}}  \def\deg{\ifm{^\circ}}

\begin{document}
%\linenumbers
\begin{frontmatter}
\title{\mathversion{bold} Measurement of the running of the fine structure constant below 1 GeV with the KLOE Detector}

\collab{The KLOE-2 Collaboration}
\vspace{-0.4cm}
\author[Messina,Frascati]{A.~Anastasi},
%\author[Roma2,INFNRoma2]{F.~Archilli},
\author[Frascati]{D.~Babusci},
%\author[Roma2,INFNRoma2]{D.~Badoni},
%\author[Cracow]{I.~Balwierz-Pytko},
\author[Frascati]{G.~Bencivenni},
\author[Warsaw]{M.~Berlowski},
%\author[Roma1,INFNRoma1]{C.~Bini},
\author[Frascati]{C.~Bloise},
%\author[INFNRoma1]{V.~Bocci},
\author[Frascati]{F.~Bossi},
\author[INFNRoma3]{P.~Branchini},
\author[Roma3,INFNRoma3]{A.~Budano},
%\author[Moscow]{S.~A.~Bulychjev}
\author[Uppsala]{L.~Caldeira~Balkest\aa hl},
\author[Uppsala]{B.~Cao},
%\author[Frascati]{P.~Campana},
%\author[Frascati]{G.~Capon},
\author[Roma3,INFNRoma3]{F.~Ceradini},
\author[Frascati]{P.~Ciambrone},
\author[Messina,INFNCatania,Novosibirsk]{F.~Curciarello},
%\ead{fcurciarello@unime.it}
\author[Cracow]{E.~Czerwi\'nski},
\author[Roma1,INFNRoma1]{G.~D'Agostini},
\author[Frascati]{E.~Dan\'e},
\author[INFNRoma3]{V.~De~Leo\corauthref{cor}},
\ead{veronica.deleo@roma3.infn.it}
\author[Frascati]{E.~De~Lucia},
%\author[INFNBari]{G.~De~Robertis},
%\author[Roma1,INFNRoma1]{A.~De~Santis},
\author[Frascati]{A.~De~Santis},
%\author[Roma1,INFNRoma1]{G.~De~Zorzi},
\author[Frascati]{P.~De~Simone},
\author[Roma3,INFNRoma3]{A.~Di~Cicco},
\author[Roma1,INFNRoma1]{A.~Di~Domenico},
%\author[Napoli,INFNNapoli]{C.~Di~Donato},
\author[INFNRoma2]{R.~Di~Salvo},
%\author[Roma3,INFNRoma3]{B.~Di~Micco},
\author[Frascati]{D.~Domenici},
\author[Frascati]{A.~D'Uffizi},
%\author[Bari,INFNBari]{O.~Erriquez},
%\author[Bari,INFNBari]{G.~Fanizzi},
\author[Roma2,INFNRoma2]{A.~Fantini},
\author[Frascati]{G.~Felici},
\author[ENEACasaccia,INFNRoma1]{S.~Fiore},
%\author[Roma1,INFNRoma1]{P.~Franzini},
\author[Cracow]{A.~Gajos},
\author[Roma1,INFNRoma1]{P.~Gauzzi},
\author[Messina,INFNCatania]{G.~Giardina},
\author[Frascati]{S.~Giovannella},
%\author[Roma2,INFNRoma2]{F.~Gonnella},
\author[INFNRoma3]{E.~Graziani},
%\author[Cracow]{A.~Gruntowski}
\author[Frascati]{F.~Happacher},
\author[Uppsala]{L.~Heijkenskj\"old},
%\author[Uppsala]{B.~H\"oistad},
%\author[Frascati]{L.~Iafolla},
%\author[Energetica,Frascati]{E.~Iarocci},
%\author[Uppsala]{W.~Ikegami Andersson}
%\author[Uppsala]{M.~Jacewicz},
%\author[Berlin]{F.~Jegerlehner}
\author[Uppsala]{T.~Johansson},
%\author[Cracow]{K.~Kacprzak},
\author[Cracow]{D.~Kami\'nska},
\author[Warsaw]{W.~Krzemien},
%\author[Warsaw]{A.~Kowalewska},
%\author[Moscow]{V.~Kulikov},
\author[Uppsala]{A.~Kupsc},
%\author[Frascati,StonyBrook]{J.~Lee-Franzini},
%\author[Frascati]{B.~Leverington},
%\author[INFNBari]{F.~Loddo},
\author[Roma3,INFNRoma3]{S.~Loffredo},
\author[Novosibirsk1,Novosibirsk2]{P.A.~Lukin},
\author[Messina2,INFNMessina]{G.~Mandaglio},
%\ead{gmandaglio@unime.it}
%\author[Moscow]{M.~Martemianov},
\author[Frascati,Marconi]{M.~Martini},
\author[Frascati]{M.~Mascolo},
%\author[Moscow]{M.~Matsyuk},
\author[Roma2,INFNRoma2]{R.~Messi},
\author[Frascati]{S.~Miscetti},
\author[Frascati]{G.~Morello},
\author[INFNRoma2]{D.~Moricciani},
\author[Cracow]{P.~Moskal},
%\author[INFNRoma3,LIP]{F.~Nguyen},
%\author[Frascati]{L.~Quintieri},
\author[Uppsala]{M.~Papenbrock},
\author[INFNRoma3]{A.~Passeri},
\author[Energetica,INFNRoma1]{V.~Patera},
\author[Frascati]{E.~Perez~del~Rio},
%\author[Roma3,INFNRoma3]{I.~Prado~Longhi},
\author[INFNBari]{A.~Ranieri},
%\author[Mainz]{C.~F.~Redmer},
\author[Frascati]{P.~Santangelo},
\author[Frascati]{I.~Sarra},
\author[Calabria,INFNCalabria]{M.~Schioppa},
%\author[Frascati]{B.~Sciascia},
%\author[Energetica,Frascati]{A.~Sciubba},
\author[Roma3]{A.~Selce},
\author[Frascati]{M.~Silarski},
\author[Frascati]{F.~Sirghi},
%\author[Calabria,INFNCalabria]{S.~Stucci},
%\author[Roma3,INFNRoma3]{C.~Taccini},
\author[INFNRoma3]{L.~Tortora},
\author[Frascati]{G.~Venanzoni\corauthref{cor}},
\ead{graziano.venanzoni@lnf.infn.it}
%\author[Frascati,CERN]{R.~Versaci},
\author[Warsaw]{W.~Wi\'slicki},
\author[Uppsala]{M.~Wolke}\\
%\author[Cracow]{J.~Zdebik}
\vspace{-0.1cm}
\begin{center}
and\\
\hspace{0.5cm}\author[Berlin,Desy]{F.~Jegerlehner}
\end{center}

%%%%%%%%%%%%%%%%%%%%%%%%%%%%%%%%%%%%%%%%%%%%%%%%%%%%%%%%%%%%%%%%%%%%%%%%%%%%%%%%%%%%%%%%%%%%%%%%%%
%\address[Bari]{\affuni{di Bari}{Bari}}
\address[INFNBari]{\affinfn{Bari}{Bari}}
%\address[CentroCatania]{Centro Siciliano di Fisica Nucleare e Struttura della Materia, Catania, Italy.}
\address[INFNCatania]{\affinfn{Catania}{Catania}}
\address[Cracow]{Institute of Physics, Jagiellonian University, Cracow, Poland.}
\address[Frascati]{Laboratori Nazionali di Frascati dell'INFN, Frascati, Italy.}
%\address[Messina]{\affuni{di Messina}{Messina}}
%\address[Mainz]{Institut f\"ur Kernphysik,
%Johannes Gutenberg Universit\"at Mainz, Germany.}
\address[Messina]{Dipartimento di Scienze Matematiche e Informatiche, Scienze Fisiche e Scienze della Terra dell'Universit\`a di Messina, Messina, Italy.}
\address[Messina2]{Dipartimento di Scienze Chimiche, Biologiche, Farmaceutiche ed Ambientali dell'Universit\`a di Messina, Messina, Italy.}
\address[INFNMessina]{INFN Gruppo collegato di Messina, Messina, Italy.}
\address[Calabria]{\affuni{della Calabria}{Rende}}
\address[INFNCalabria]{INFN Gruppo collegato di Cosenza, Rende, Italy.}
%
%\address[Moscow]{Institute for Theoretical and Experimental Physics (ITEP), Moscow, Russia.}
%\address[Napoli]{\affuni{``Federico II''}{Napoli}}
%\address[INFNNapoli]{\affinfn{Napoli}{Napoli}}
\address[Energetica]{Dipartimento di Scienze di Base ed Applicate per l'Ingegneria dell'Universit\`a
``Sapienza'', Roma, Italy.}
\address[Marconi]{Dipartimento di Scienze e Tecnologie applicate, Universit\`a ``Guglielmo Marconi", Roma, Italy.}
\address[Novosibirsk]{Novosibirsk State University, 630090 Novosibirsk, Russia.}
\address[Roma1]{\affuni{``Sapienza''}{Roma}}
\address[INFNRoma1]{\affinfn{Roma}{Roma}}
\address[Roma2]{\affuni{``Tor Vergata''}{Roma}}
\address[INFNRoma2]{\affinfn{Roma Tor Vergata}{Roma}}
\address[Roma3]{Dipartimento di Matematica e Fisica dell'Universit\`a
``Roma Tre'', Roma, Italy.}
%\address[Roma3]{\affuni{``Roma Tre''}{Roma}}
\address[INFNRoma3]{\affinfn{Roma Tre}{Roma}}
\address[ENEACasaccia]{ENEA UTTMAT-IRR, Casaccia R.C., Roma, Italy}
%\address[Berlin]{Institute of Physics Humboldt-University of Berlin, Berlin, Germany}
\address[Uppsala]{Department of Physics and Astronomy, Uppsala University, Uppsala, Sweden.}
\address[Warsaw]{National Centre for Nuclear Research, Warsaw, Poland.}
\address[Berlin]{Institute of Physics Humboldt-University of Berlin, Berlin, Germany}
\address[Desy]{Deutsches  Elektronen--Synchrotron (DESY), Platanenallee 6, D--15738 Zeuthen, Germany}
\address[Novosibirsk1]{Budker Institute of Nuclear Physics, Novosibirsk, 630090, Russia}
\address[Novosibirsk2]{Novosibirsk State University, Novosibirsk, 630090, Russia}
%\address[CERN]{Present Address: CERN, CH-1211 Geneva 23, Switzerland.}
%\address[LIP]{Present Address: Laborat\'orio de Instrumenta\c{c}\~{a}o e F\'isica Experimental de Part\'iculas,
%Lisbon, Portugal.}
\corauth[cor]{Corresponding authors}

\begin{abstract}
We have measured the running of the effective QED coupling constant $\alpha(s)$ in the time-like region $0.6<\sqrt s< 0.975$ GeV with the KLOE detector at DA$\Phi$NE using the Initial-State Radiation process $e^+e^-\to\mu^+ \mu^-\gamma$.
%normalized to large angle Bhabha scattering. 
It represents the first measurement of the running of $\alpha(s)$ in this energy region. 
%The results show a 5$\sigma$ evidence of the hadronic contribution to the running of $\alpha_{QED}(s)$. 
Our results show a more than 5$\sigma$ significance of the hadronic contribution to the running of $\alpha(s)$, which is the %strongest direct evidence ever achieved both 
%in time- and space-like regions
% by a single experiment.
strongest direct evidence both in time- and space-like regions achieved in a single measurement. 
By using the $e^+e^-\to\pi^+\pi^-$% cross section previously measured at KLOE, 
cross section measured by KLOE,
the 
%$2\pi$ contribution to the 
real and imaginary parts of the shift $\Delta\alpha(s)$ have been extracted.
From a fit of the real part of $\Delta\alpha(s)$
% the product of the Branching fractions $BR(\omega\to\mu^+\mu^-)BR(\omega\to e^+e^-)$ has been measured.
and assuming the lepton universality the branching ratio $BR(\omega\to\mu^+\mu^-) = (6.6\pm1.4_{stat}\pm1.7_{syst})\cdot 10^{-5} $ has been determined.
\end{abstract}
\end{frontmatter}
\clearpage
%\begin{linenumbers}
\def\spreadlines#1{\par\renewcommand\baselinestretch{#1}\normalsize}
\def\red{\color{red}}  \def\blue{\color{blue}}

\section{Introduction}\label{Introduction}
Precision tests of the Standard Model (SM) require an appropriate inclusion of higher-order effects and the  very precise knowledge of input parameters~\cite{Fred_2003}. One of the basic input parameters is the effective QED coupling constant $\alpha$, 
determined from the anomalous magnetic moment of the electron 
with the impressive accuracy of 0.37 parts per billion~\cite{Aoyama}. 
%However, physics at nonzero squared momentum transfer $q^2$ requires an effective electromagnetic coupling $\alpha(q^2)$. 
However, physics at non-zero momentum transfer requires an effective electromagnetic coupling $\alpha(s)$~\footnote{In the following we will indicate with $s$ 
%instead of $q^2$ 
the momentum transfer squared of the reaction.}. 
The shift of the fine-structure constant from
the Thomson limit to high energy involves low energy 
non-perturbative hadronic effects which affect the precision. 
These effects represent the largest uncertainty (and the main limitation) for the electroweak 
precision tests  as the determination of $sin^2\theta_W$ at the Z pole or the SM prediction of the muon $g-2$~\cite{fred}.\\
%precision tests at the Z pole (like in the electroweak (EW) mixing parameter sin$^2\theta$, related to 
%$\alpha$, the Fermi coupling constant $G_F$ and $M_{Z}$ via the Sirlin relation~\cite{Sirlin}), and in the SM prediction of the muon $g-2$~\cite{fred}.\\
The QED coupling constant is predicted
%~\cite{Arbuzov:2004wp}
and observed~\cite{Arbuzov:2004wp,expalpha} to increase with rising momentum transfer
%~\footnote{This is different from the strong coupling constant $\alpha_S$ which decreases with rising $q^2$.},
(differently from the strong coupling constant $\alpha_S$ which decreases with rising momentum transfer),
 which 
%The strengthening of the QED coupling with rising momentum 
can be understood as a result of the screening of the 
 bare charge caused by the polarized cloud of virtual particles.
% a polarization of the vacuum. 
%The observed charge is therefore reduced due to the screening of the virtual particles.\\
%\\
%Thus, at large distances, the observed bare charge is reduced, due to this effect of screening. %As we probe closer we penetrate 
%As we probe closer we enter
%into the cloud of virtual particles, decreasing the screening effect and observing a larger fraction of the bare charge and thus a strengthening
%of the coupling. \\  
The vacuum polarization (VP) effects can be absorbed 
in a redefinition of the fine-structure constant, making it $s$ dependent:
%The effective  fine-structure constant at squared momentum transfer $q^2$ 
% is usually defined as: \alpha(q^2) = {{\alpha(0)} \over {1-\Delta \alpha(s)}}
%\label{running}
\begin{equation}
\alpha(s) = {{\alpha(0)} \over {1-\Delta \alpha(s)}}.
\label{running}
\end{equation}

The shift $\Delta\alpha$(s) in terms of the vacuum polarization function $\Pi'_\gamma$(s) is given by:
\begin{equation}
 \Delta \alpha(s) = -4\pi\alpha(0)\,\rm Re\,[\Pi_\gamma^{'}(s)-\Pi_\gamma^{'}(0)]
\label{delta_alpha}
\end{equation}
and it is the sum of the lepton ($e$,$\mu$,$\tau$) contributions, the 
contribution from the five quark flavours (u,d,s,c,b), and the contribution of the top quark (which can be neglected at low energies): $\Delta\alpha$(s)=$\Delta\alpha_{lep}(s)+
\Delta\alpha_{had}^{(5)}(s)+\Delta\alpha_{top}(s)$\cite{Fred_2003}.\\
The leptonic contributions can be calculated with very high precision in QED using the perturbation theory \cite{passera,sturm}.
However, due to the non-perturbative behaviour of the strong
interaction at low energies, perturbative QCD only allows
us to calculate the high energy tail of the hadronic (quark)
contributions. In the lower energy region the hadronic contribution can be evaluated through a dispersion integral over the measured $e^+e^- \rightarrow$ hadrons cross-section:

\begin{equation}
\Delta\alpha_{had}(s)=-\big(\frac{\alpha(0) s}{3\pi}\big){\rm Re}\int_{ 4m_\pi^2}^\infty ds'\, \frac{R_{had}(s')}{s'(s'-s-i\epsilon)} ,
\end{equation}
where $R_{had}(s)$ is defined as the cross section ratio $R_{had}(s)=\frac{\sigma(e^+e^-\rightarrow\gamma*\rightarrow hadrons)}{\sigma(e^+e^-\rightarrow\gamma*\rightarrow \mu^+\mu^-)}$.

%The main difficulty of this approach is that experimental data exhibit errors, which give the dominant uncertainty in the evaluation of $\Delta\alpha$ \cite{Opal}.\\
In this approach the dominant uncertainty in the evaluation of $\Delta\alpha$ is given by 
the experimental data accuracy. \\
%Equations (\ref{running}) and (\ref{delta_alpha}) are the usual definition of the running
%effective QED coupling and have the advantage that they define a real coupling.\\
In the Eq. (\ref{delta_alpha}) $\rm Im\,\Delta\alpha$ related to the imaginary part of the VP function $\Pi'_\gamma$ is completely neglected, which is a good approximation in the continuum as the contributions from the imaginary part are suppressed.
However, this approximation is not sufficient in the presence of resonances like the $\rho$ meson,
where the accuracy of the cross section measurements reaches the order of (or even less than) 1\%, and the imaginary part should be taken into account.\\
In this paper we present a measurement of the running of 
the effective QED coupling constant $\alpha$
in the time-like region 0.6$<\sqrt{s}<$0.975 GeV. The strength of the coupling constant is measured as a function of the momentum transfer of the
exchanged photon $\sqrt{s}=M_{\mu\mu}$ where $M_{\mu\mu}$ is the  $\mu^+\mu^-$ invariant mass.% by computing 
\,The value of $\alpha(s)$ is extracted from the ratio of the  differential cross section 
%$d\sigma^{ISR}/d\sqrt{s}$
 for the process $e^+e^- \rightarrow \mu^+\mu^- \gamma(\gamma)$ with the photon emitted in the Initial State (ISR) to the corresponding cross section obtained from Monte Carlo (MC) simulation 
%$d\sigma^{ISR}_{MC,0}/d\sqrt{s}$  
with the coupling set to the constant value $\alpha (s)= \alpha(0)$:
\begin{equation}
\lvert \frac{\alpha(s)}{\alpha(0)}\rvert^2= \frac{d\sigma_{data} (e^+e^- \rightarrow \mu^+\mu^- \gamma(\gamma))\vert_{ISR}/d\sqrt{s}}{d\sigma^{0}_{MC}(e^+e^- \rightarrow \mu^+\mu^- \gamma(\gamma))\vert_{ISR}/d\sqrt{s}}
\label{our_method}
\end{equation}
To obtain the ISR cross section, the observed cross section must be corrected for events with one or more photons in the final state (FSR). This has been done by using the PHOKHARA MC event generator,  which includes  next-to-leading-order ISR and FSR contributions~\cite{PHOKHARA}.
In the following we only use events where the photon is emitted at small angles, which results in a large enhancement of the ISR with respect to the FSR contribution. 
From the measurement of the effective coupling constant and the dipion cross section \cite{Venanzo}, we extracted for the first time in a single experiment
the real and imaginary part of $\Delta\alpha$.\\
The analysis has been performed by using the data collected with the KLOE detector at DA$\Phi$NE~\cite{Gallo:2006yn}, the $e^+e^-$ collider running at the $\phi$ meson mass, with a total integrated luminosity of 1.7 fb$^{-1}$.  

\section{The KLOE Detector}\label{KLOE}
The KLOE detector consists of a cylindrical drift chamber (DC)
\cite{Adinolfi1} and an electromagnetic calorimeter (EMC) \cite{Adinolfi2}. The DC has a
momentum resolution of $\sigma_{p_\perp}/p_\perp \sim 0.4\%$ for tracks with polar angle
$\theta > 45^\circ$. Track points are measured in the DC with a resolution
in $r-\phi$ of $\sim$ 0.15 mm and $\sim$ 2 mm in z. The EMC has an energy
resolution of $\sigma_E/E \sim 5.7\%/$ 
$\sqrt E(GeV)$ and an excellent time resolution
of $\sigma_t \sim 54$\,ps$/\sqrt E $\,(GeV) $\oplus$ 100 ps.
Calorimeter clusters are
reconstructed grouping together energy deposits close in space and
time. A superconducting coil provides an axial magnetic field of
0.52 T along the bisector of the colliding beam directions. The bisector
is taken as the z axis of our coordinate system. 
The x axis is horizontal and the y
axis is vertical, directed upwards. A cross section of the detector in
the y, z plane is shown in Fig.\ref{kloe_section}. 
The trigger uses both EMC and DC information. Events used in this
analysis are triggered by two energy deposits larger than 50 MeV
in two sectors of the barrel calorimeter.

\subsection{Event selection}\label{Data analysis}
\begin{figure}
\centerline{
\includegraphics[width=12pc]{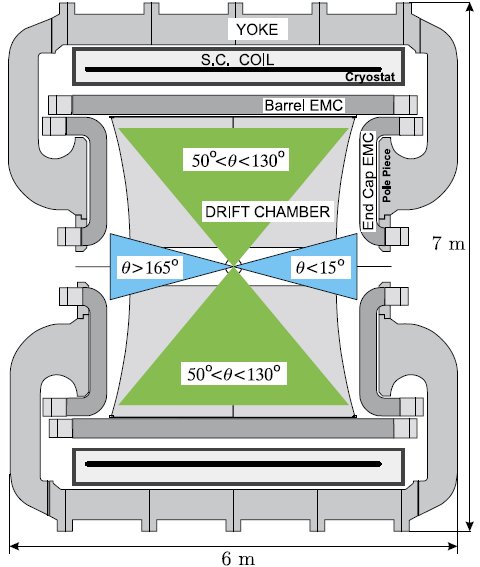}}
\caption{Detector section with the acceptance region for the charged tracks (wide cones)
and for the photon (narrow cones).}
\label{kloe_section}
\end{figure}

%The data sample consists of 1.7 fb$^{-1}$ data taken in the years 2004-2005.
A photon and two tracks of opposite curvature are required to identify a $\mu\mu\gamma$ event. 
%We used selection cuts in which the photon (undetected) is emitted 
Events are selected with a (undetected) photon emitted
 at small angle (SA),{\it i.e.} within a cone of $\theta_\gamma < 15^\circ$ around the beamline (narrow cones in Fig. \ref{kloe_section}) %and the two charged muons tracks have $^\circ<\theta_\mu<130^\circ$ 
%(wide cones in fig. \ref{kloe_section}).
and the two charged muons are emitted at large polar angle, $50^\circ<\theta_\mu<130^\circ$.
High statistics for the ISR signal and significant reduction of
background events as $\phi \rightarrow \pi^+ \pi^- \pi^0$ in which the $\pi^0$ mimics the missing momentum of the photon(s) 
and from the FSR radiation process,
$e^+ e^- \rightarrow \mu^+ \mu^- \gamma_{FSR}$, are guaranteed by this selection. 
However, this requirement 
%On the other hand this selection 
results in a kinematical suppression of events with $\sqrt{s}<$\,0.6 GeV, since a highly energetic photon emitted at small angle forces
the muons also to be at small angles (and thus outside the acceptance).\\
To avoid spiralling tracks in the drift chamber, 
the reconstructed momenta must have $p_T>$ 160
MeV or $\vert p_z \vert >$ 90 MeV. 
This ensures good reconstruction and efficiency. \\
%qui
The main background reactions are given by: 
\begin{itemize}
\item $e^+e^-\rightarrow \pi^+\pi^-\gamma(\gamma)$
\item $e^+e^-\rightarrow \pi^+\pi^-\pi^0$
\item $e^+e^-\rightarrow e^+e^- \gamma(\gamma)$.
\end{itemize}

\begin{figure}[h!]
\centerline{
\includegraphics[width=71.8mm]{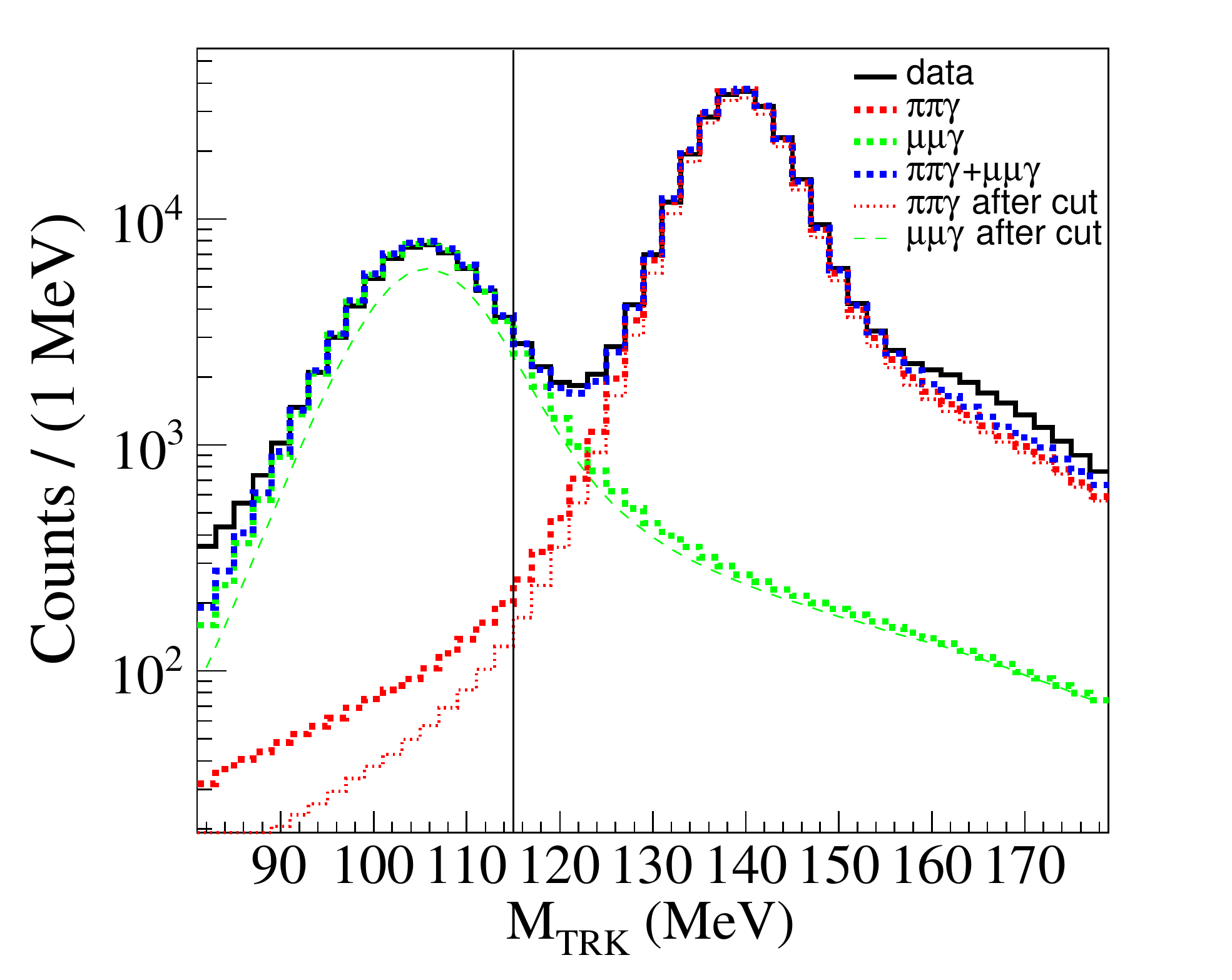}}
\caption{$\pi\pi\gamma$ and $\mu\mu\gamma$ $M_{TRK}$ distributions. The vertical line shows the $\mu\mu\gamma$ selection cut ($M_{TRK}<115$ MeV). The effect of the $\sigma_{M_{TRK}}$ cut on the two distributions is clearly visible.}
\label{smtrk}
\end{figure}

A particle ID estimator (PID) based on a pseudo-likelihood function (L$_\pm$) using time-of-flight and calorimeter information (size and shape of the energy deposit) is used to obtain separation between electrons and pions or muons.
Events with both tracks satisfying L$_\pm<$0 are rejected as $e^+e^-\gamma$.
To separate the muons from the pions we applied mainly two cuts: the first on the track mass ($M_{TRK}$) variable and the second  on the $\sigma_{MTRK}$, the estimated error on $M_{TRK}$. %(shown in \ref{smtrk}). 
Assuming the
presence of only one unobserved photon and that the tracks belong to particles of
the same mass, $M_{TRK}$ is computed from energy and momentum conservation.
The $\sigma_{MTRK}$ variable is constructed event by event with the error matrix of the 
fitted tracks at the point of closest approach (PCA) \cite{Fra}. 
Cutting the high values of this variable the bad reconstructed tracks are rejected allowing a reduction of the $\pi\pi\gamma$ events contamination (shown in Fig.~\ref{smtrk}). 

Residual %$\pi^+\pi^-\gamma$, $\pi^+\pi^-\pi^0$ and $e^+e^-\gamma$ 
background is evaluated by fitting the observed   $M_{TRK}$ spectrum with a superposition
of MC simulation distributions describing signal
and $\pi^+\pi^-\gamma$, $\pi^+\pi^-\pi^0$ and $e^+e^-\gamma$ events.
The normalization factors from signal and backgrounds are free parameters of the
fit, performed for 30 intervals in $s$ of 0.02 GeV$^2$ width for 0.35$<$\textit{s}$<$0.95 GeV$^2$.
Additional background from the $e^+e^- \rightarrow e^+e^-\mu^+\mu^-$ process has 
been evaluated using the NEXTCALIBUR MC generator~\cite{nextcalibur}. 
%It contributes 
%as maximum up to 
The maximum contribution is
0.7\% at $\sqrt{s}$=0.6\,GeV. The uncertainty on this background
has been taken as 50\% of the total contribution and 
added to the systematic error. The contribution from $e^+e^- \rightarrow e^+e^-\pi^+\pi^-$ has been evaluated with the EKHARA generator\cite{ekhara} and found to be negligible.\\ 
The total fractional systematic uncertainty on background subtraction, obtained by adding in quadrature the uncertainties on the fit normalization parameters and  the $e^+ e^-\mu^+\mu^-$ residual background, 
ranges from 0.2\% to 0.05\% decreasing with $s$.\\
About $4.5\cdot 10^6$  $\mu\mu\gamma$ events pass these selection criteria.

\section{Measurement of the $\mu\mu\gamma$ cross section}
\begin{figure}[h!]
\begin{center}
\includegraphics[width=10cm]{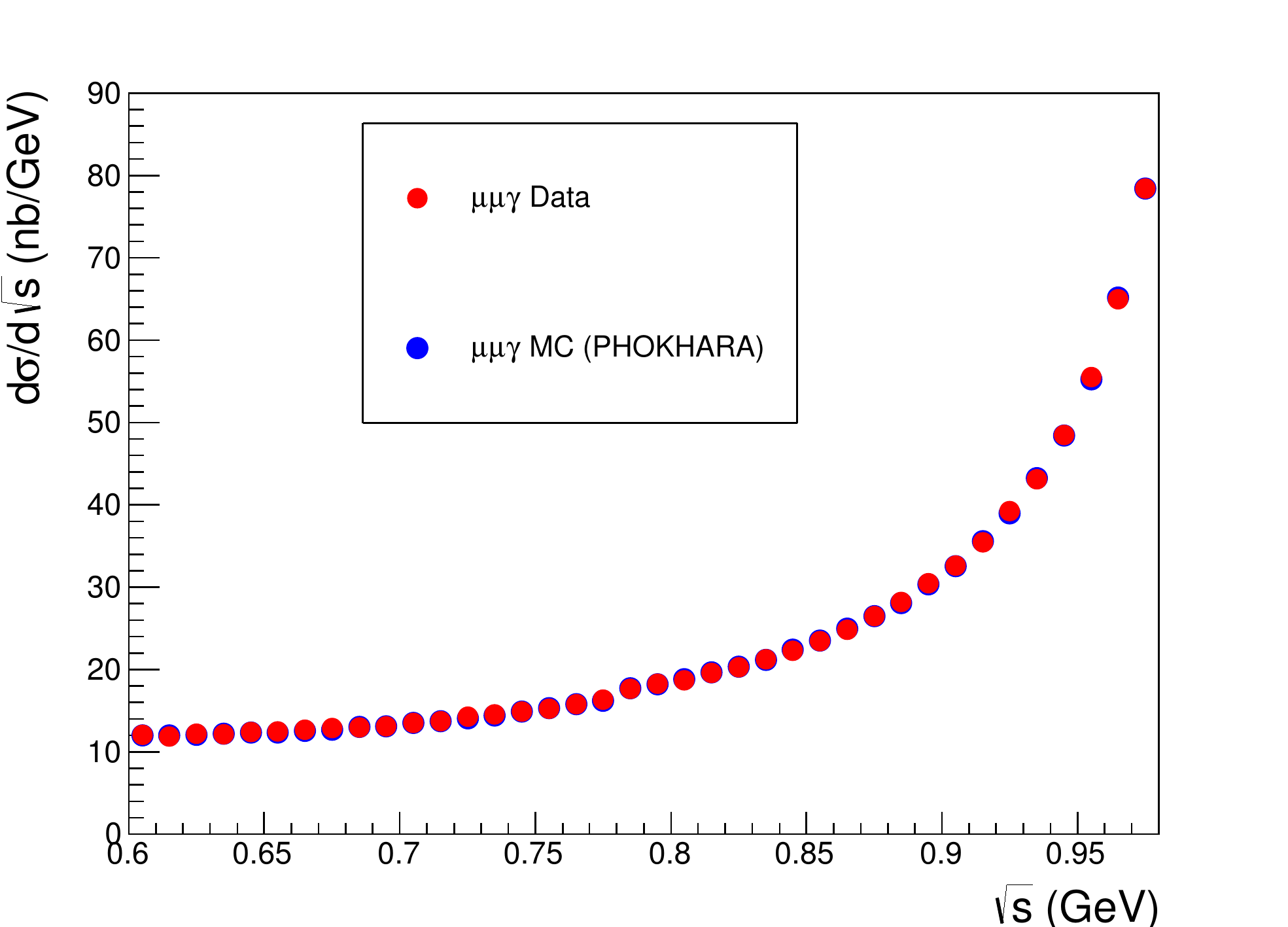}
\includegraphics[width=10cm]{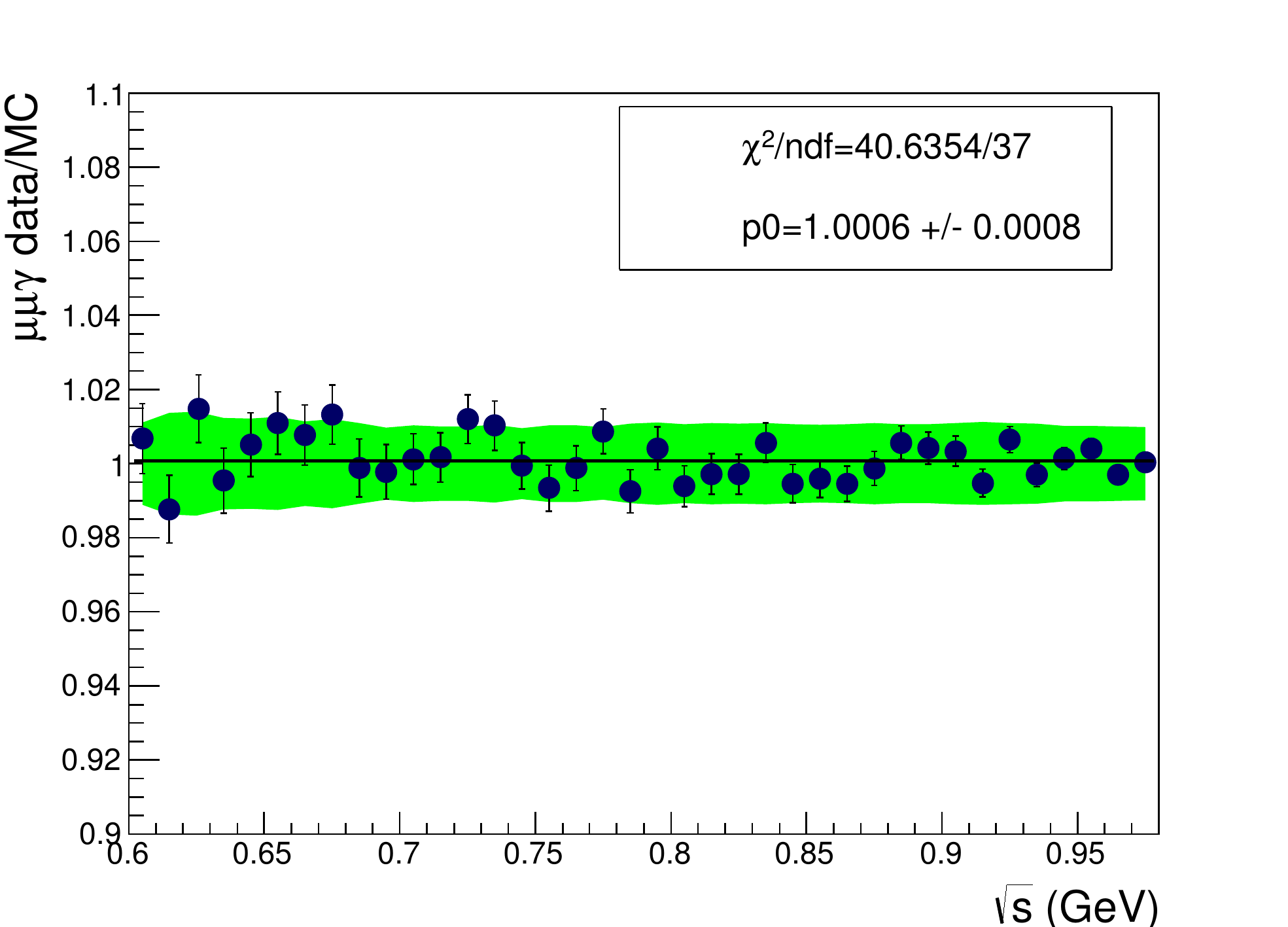}
\caption{Upper plot: comparison of the measured differential cross section  (red points) and PHOKHARA
MC prediction (blue points) of the $\mu^+\mu^-\gamma$ cross section. Lower plot: the
ratio of the two. The green band shows the systematic error.}
%}
\label{mmg_abs_q.eps}
\end{center}
\end{figure}

The experimental ISR $\mu^+\mu^-\gamma$ cross section is obtained from the observed number of events ($N_{obs}$) and the background estimate ($N_{bckg}$) as: 

\begin{equation}
\frac{d\sigma(e^+e^-\rightarrow \mu^+\mu^-\gamma(\gamma))}{d\sqrt{s}}\biggr\rvert_{ISR}= \frac{N_{obs}-N_{bkg}}{\Delta\sqrt{s}}\cdot\frac{(1-\delta_{FSR}) }{\epsilon(\sqrt{s})\cdot \textit{L}},
\label{mmg}
\end{equation}

where $(1-\delta_{FSR})$ is the correction applied to remove the FSR contribution (which increases with the energy from 0.998 at 0.605 GeV to 1.032 at 0.975 GeV), 
%(shown in fig.\ref{fsr_correction}),
$\epsilon$ is the efficiency (see section below)  and $L$ is the integrated luminosity. \\
We firstly compare the $\mu^+\mu^-\gamma$ cross-section with only ISR 
 with the corresponding NLO QED calculation from PHOKHARA generator including the VP effects.\\
In the upper plot of Fig. \ref{mmg_abs_q.eps} the measured $\mu^+\mu^-\gamma$ cross-section as a  function of $\sqrt{s}$ for both experimental (red points) and MC (blue points) data is shown.  
The agreement between the two cross sections is excellent. 
The same figure shows an interesting feature around 0.78 GeV (corresponding to the  mass of the $\omega$ meson), where a small step appears in the cross section. This step behaviour  is due to the $\rho-\omega$ interference in the photon propagator, as it will be shown in the following.
In the lower plot the data to MC ratio is shown together with the systematic error (green band) of the order of 1\%.\\

\section{Efficiencies and systematic errors}
The global efficiency,  which ranges from 0.086 at 0.605 GeV to 0.27 at 0.975 GeV,
 has been obtained from a $\mu^+ \mu^- \gamma$ events generation with PHOKHARA interfaced with the detector simulation code GEANFI\cite{geanfi}. It includes contributions from trigger, tracking, PID, $\sigma_{MTRK}$, $M_{TRK}$ and acceptance.\\
\noindent\textbf{Trigger}: the trigger efficiency has been obtained from a sample of $\mu^+\mu^-\gamma$ events where a single muon satisfies the trigger requirement. Trigger response for the other muon is parameterized as a function of its momentum and direction. The efficiency as a function of $s$  is obtained using the MC event distribution and differs from one by less than $10^{-4}$, with negligible systematic error.\\
\textbf{Tracking}: the single muon track efficiency has been obtained as a function of the particle momentum and polar angle by means of a high purity $\mu^+ \mu^- \gamma$ sample obtained by using one muon to tag the presence of the other.
The combined efficiency is about 99\%, almost constant in $s$. The systematic uncertainty on tracking efficiency is evaluated changing the purity of the control sample and ranges from $0.3$ to $0.6\%$ as a function of $s$. \\
\textbf{PID}: The signal efficiency due to
the PID cut is more than 99.5\%, as evaluated with $\mu^+\mu^-\gamma$ samples obtained both from data and Monte Carlo, with negligible systematic error.\\
\textbf{\bf$M_{\rm \bf TRK}$, \bf\mathversion{bold}$\sigma_{M_{\mathrm{TRK}}}$ and Acceptance cuts}: 
Efficiencies are taken from MC, with systematic errors obtained as:
\begin{itemize}
\item The systematic uncertainty due to $M_{TRK}$ cut has been obtained by varying the cut by one standard deviation of the mass resolution and evaluating
the difference in the $\mu\mu\gamma$ spectrum. We find a fractional
difference of 0.4\% (constant in $s$) which we take as systematic error.\\
\item The systematic uncertainty on $\sigma_{M_{\mathrm{TRK}}}$ cut has been evaluated as the maximum difference between the $\mu \mu \gamma$ normalization parameters of the background fitting procedure, obtained with standard cuts, and those obtained by shifting $\sigma_{M_{\mathrm{TRK}}}$ by $\pm$2\%.
The contribution is less than 1\% in the whole energy range.\\
\item Systematic effects due to polar angle requirements for the
muons and for the photon, are estimated by varying the angular acceptance by 
$\pm 1^\circ$ (more than two times the resolution on the polar angle) around
the nominal value. The uncertainty ranges from 0.1 to 0.6\%.\\
\end{itemize}
\textbf{Software trigger}: A third-level trigger is implemented 
 to keep the physics events which are misidentified as cosmic rays. 
Its efficiency for $\mu\mu\gamma$
events, evaluated from an unbiased downscaled sample, is consistent
with one within 10$^{-3}$ which is taken as systematic error.\\
Table~\ref{tab:syseff1} gives the systematic errors at the $\rho$-peak mass value. 
\section{Luminosity and Radiative corrections}
Large angle Bhabha scattering is used to determine the luminosity, with a reference cross section obtained with Babayaga@NLO MC event generator~\cite{babayaga}, convolved with detector and beam conditions~\cite{lumi}. Two sources contribute to the systematic uncertainty in the evaluation of the luminosity:
\begin{itemize}
\item the theoretical accuracy of Babayaga@NLO, quoted as 0.1\% by the authors;
\item the systematic error associated to the counting of Bhabha events which is 0.3\%~\cite{lumi}
%\item the dependence of the Bhabha cross section on the VP effect.
\end{itemize}

When extracting the running of $\alpha$ (see following Section), the dependence of the Bhabha cross section on the VP effect must be taken into account.
By switching off the hadronic corrections to the VP, we checked that the presence of the hadronic contribution to $\Delta\alpha$ for both $s$ and $t$ channels 
%($\Delta\alpha_{had}(s,t)$)
 in the cross section gives a 0.2\% contribution which we consider as a systematic error of our measurement ({\it $\Delta\alpha_{had}$ dep.} in Table~\ref{tab:syseff1}).
\\
The uncertainty on PHOKHARA MC generator ({\it Rad. function H} in Table~\ref{tab:syseff1}) is 0.5\% constant in $s$, mostly due to missing ISR higher-order terms\cite{PHOKHARA}.
The uncertainty in the procedure to subtract the FSR contribution is 0.2\%, mostly due to missing FSR diagrams~\cite{Campanario:2013uea}.

\begin{table}
\begin{center}
\begin{tabular}{||l|c|c||}
\hline\hline
Source &  $\sigma_{\mu\mu\gamma}$ & $|\alpha(s)/\alpha(0)|^2$ \\
\hline\hline
Trigger & \multicolumn{2}{|c||}{$<$ 0.1\%} \\
Tracking  &  \multicolumn{2}{|c||}{$s$ dep. (0.5\% at $\rho$-peak)} \\
Particle ID  & \multicolumn{2}{|c||}{$<$ 0.1\%} \\
Background subtraction & \multicolumn{2}{|c||}{$s$ dep. (0.1\% at $\rho$-peak)} \\
$M_{TRK}$  & \multicolumn{2}{|c||}{0.4\%} \\
$\sigma_{MTRK}$ & \multicolumn{2}{|c||}{$s$ dep. ($<$ 0.1\% at $\rho$-peak)} \\
Acceptance &  \multicolumn{2}{|c||}{$s$ dep. (0.3\% at $\rho$-peak)}  \\
Software Trigger & \multicolumn{2}{|c||}{0.1\%} \\
Luminosity & \multicolumn{2}{|c||}{0.3\%}\\
$\Delta\alpha_{had}$  dep. (Normalization)  & - & 0.2\%\\
FSR treatment &\multicolumn{2}{|c||}{0.2\%} \\
Rad. function $H$  & - & 0.5\% \\
%\hline
%Total theory systematics & 0.2 \\
\hline\hline
Total systematic error & $s$ dep. (0.8\% at $\rho-$peak)& (1\% at $\rho-$peak)  \\
\hline\hline
\end{tabular}
\vspace{0.2cm}
\caption{List of systematic errors.}
\label{tab:syseff1}
\end{center}
\end{table}

\section{\mathversion{bold} Measurement of the running of $\alpha$}
\label{VP}
\begin{figure}[h]
\begin{center}
\includegraphics[width=12cm]{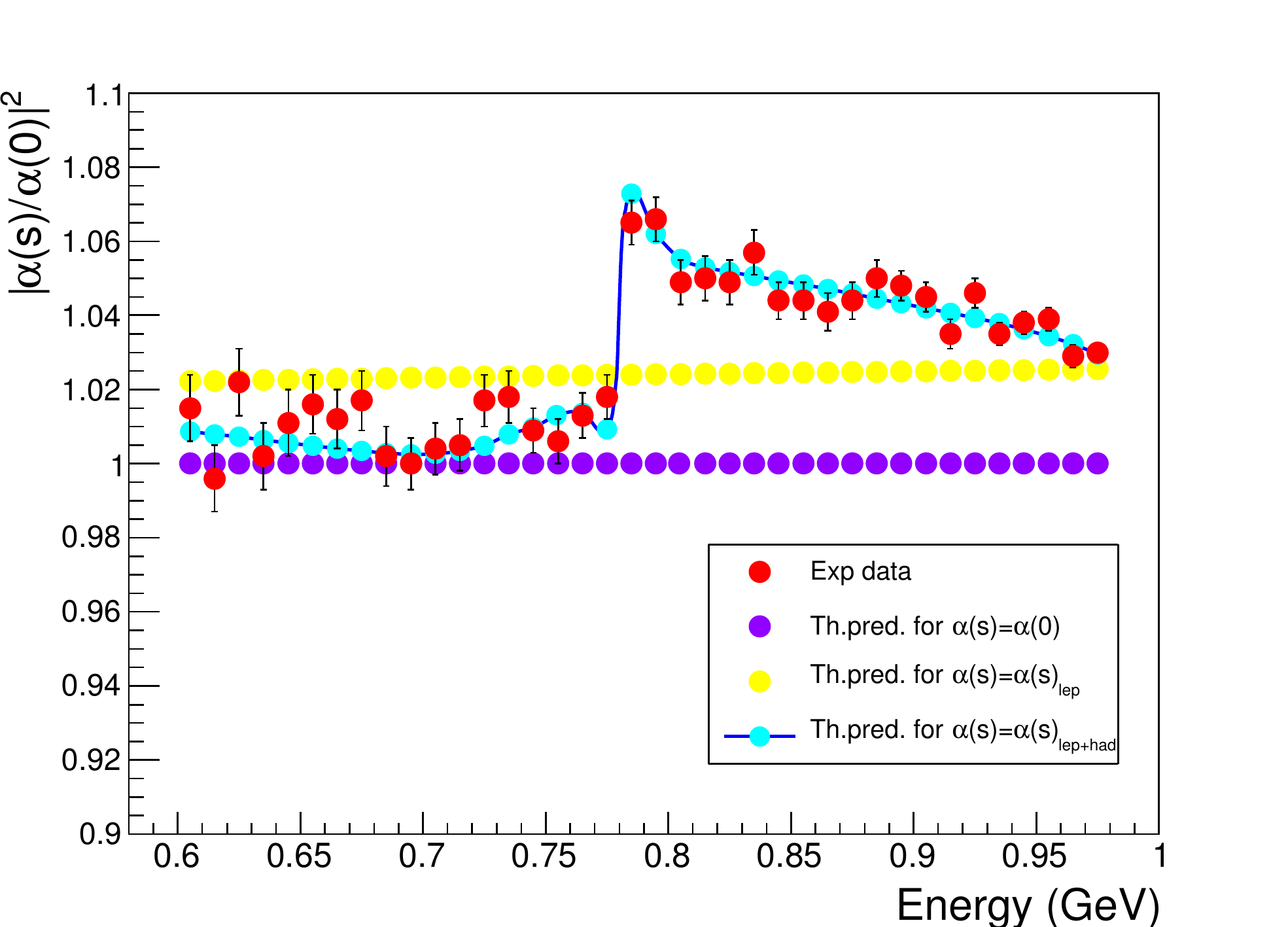}
\caption{The square of the modulus of the running $\alpha(s)$ in units of $\alpha(0)$ compared with the %``theoretical
prediction (provided by the {\tt alphaQED} package \cite{fj}) 
as a function of the dimuon invariant mass. The red points are the 
KLOE data with statistical errors; the violet points are the theoretical prediction for a fixed coupling ($\alpha(s)$ = $\alpha(0)$);
the yellow points are the prediction with only virtual lepton pairs contributing to the shift $\Delta\alpha(s)$ = $\Delta\alpha(s)_{lep}$, 
and finally the points with the solid line are the full QED prediction with both lepton and quark pairs contributing to the shift $\Delta\alpha(s)$ = $\Delta\alpha(s)_{lep+had}$.
}
\label{vp}
\end{center}
\end{figure}

We use  Eq. (\ref{our_method}) and Eq. (\ref{mmg}) in the angular region 
$\theta_\gamma < 15^\circ$
to extract the running of the effective QED coupling constant $\alpha(s)$. 
%By 'undressing' the MC cross section for VP effects ({\it i.e. by excluding these effects}), the hadronic contribution to the photon propagator,
By setting in the MC the electromagnetic coupling to the constant value $\alpha(s)$ = $\alpha(0)$, the hadronic contribution to the photon propagator,
 with its
characteristic
 $\rho-\omega$ interference structure, is clearly visible in the data to MC ratio, as shown in Fig.~\ref{vp}. The prediction from Ref.\cite{fj}  is also shown. While the leptonic part is obtained by perturbation theory, the hadronic contribution
 to $\alpha(s)$ is
obtained via an evaluation in terms of a weighted average
compilation of $R_{had}(s)$, based on the available experimental $e^+e^- \to
\mathrm{hadrons}$ annihilation data (for an up to date compilation see\cite{Jegerlehner:2015stw} and references therein). 

For comparison, the prediction with constant coupling ({\it no running}) and with only lepton pairs contributing to the running of $\alpha(s)$ is given.\\
The value of  $|\alpha(s)/\alpha(0)|^2$ with the statistical and systematic uncertainty is reported in Table~\ref{tab_alpha} . As can be seen, the total uncertainty is at the 1\% level.\\
In order to evaluate the statistical significance of the hadronic contribution to the running of $\alpha(s)$, a $\chi^2$ based statistical test for two
 different running hypotheses: (a) {\it no running}; (b) {\it running} due to lepton pairs only is performed. 

By including statistical and systematics errors, we exclude the only-leptonic hypothesis at $6\,\sigma$
which is the strongest direct evidence ever achieved by a single experiment.
Our result is also 
consistent with the estimate of $\Delta\alpha(s)$ of Ref.\cite{ignatov}
with a %statistical significance 
$\chi^2$ probability of 0.3 ($\chi^2/ndf=41.2/37$).\\   
Similar results are obtained using different $\Delta\alpha(s)$ predictions in Ref.~\cite{teubner,ignatov}.\\

\section{Extraction of Real and Imaginary part of $\Delta\alpha(s)$}
In the contribution to the running of $\alpha$, the imaginary part is
usually neglected. This is a good
approximation as the contribution from the imaginary part of
$\Delta\alpha$ enters at order $O(\alpha^2)$ compared to $O(\alpha)$ for
the real part, and is suppressed~\cite{actis}.
However, the imaginary part should be taken into account 
in the presence of resonances like the $\rho$ meson, where the cross section is measured with an accuracy better than 1\%.\\
By using the definition of the running of $\alpha$ (Eq. (\ref{running}))
the real part of the shift $\Delta\alpha(s)$ can be expressed in terms of its imaginary part and $|\alpha(s)/\alpha(0)|^2$:
%as in the following:
\begin{equation}
\rm Re\,\Delta \alpha = 1-\sqrt{\vert \alpha(0)/\alpha(s) \vert^2-(\rm Im\,\Delta \alpha)^2}.
\label{re_delta_alpha}
\end{equation}  
The imaginary part of $\Delta\alpha(s)$ can be related to the 
total cross section $\sigma(e^+e^-\to\gamma^*\to anything)$,
  where the precise relation reads~\cite{fred,nyffeler,simon&fred}:
 \vspace{0.5cm}
${\rm Im}\,\Delta \alpha = - \frac{\alpha}{3}\,R(s)$,
with $R(s) = \sigma_{tot}/\frac{4\pi\vert\alpha(s)\vert^2}{3s}$.
%\begin{equation}
%R(s) = \sigma_{tot}/\frac{4\pi\alpha(s)^2}{3s}
%\end{equation}
$R(s)$ takes into account leptonic and hadronic contribution
$R(s)=R_{lep}(s)+R_{had}(s)$, %where the leptonic part reads:
where the leptonic part corresponds to the production
of a lepton pair at lowest order taking into account mass effects:

\begin{equation}
R_{lep}(s)=\sqrt{1-\frac{4m_l^2}{s}} \left(1+\frac{2m_l^2}{s}\right), \;\; (l=e,\mu,\tau). 
\end{equation}

In the energy region around the $\rho$-meson we can approximate the hadronic cross section by the 2$\pi$ dominant contribution:

\begin{equation}
R_{had}(s)= \frac{1}{4} \left(1-\frac{4m_\pi^2}{s} \right)^\frac{3}{2} \vert F_\pi^0(s) \vert ^2,
\end{equation}

where $F_\pi^0$ is the pion form factor deconvolved:
$\lvert F_\pi^0(s)\rvert ^2 = \lvert F_\pi(s) \rvert ^2\left\lvert \frac{\alpha(0)}{\alpha(s)}\right\rvert^2$.

\vspace{0.5cm}

\begin{figure}[htp!]
\begin{center}
\includegraphics[width=10.8cm]{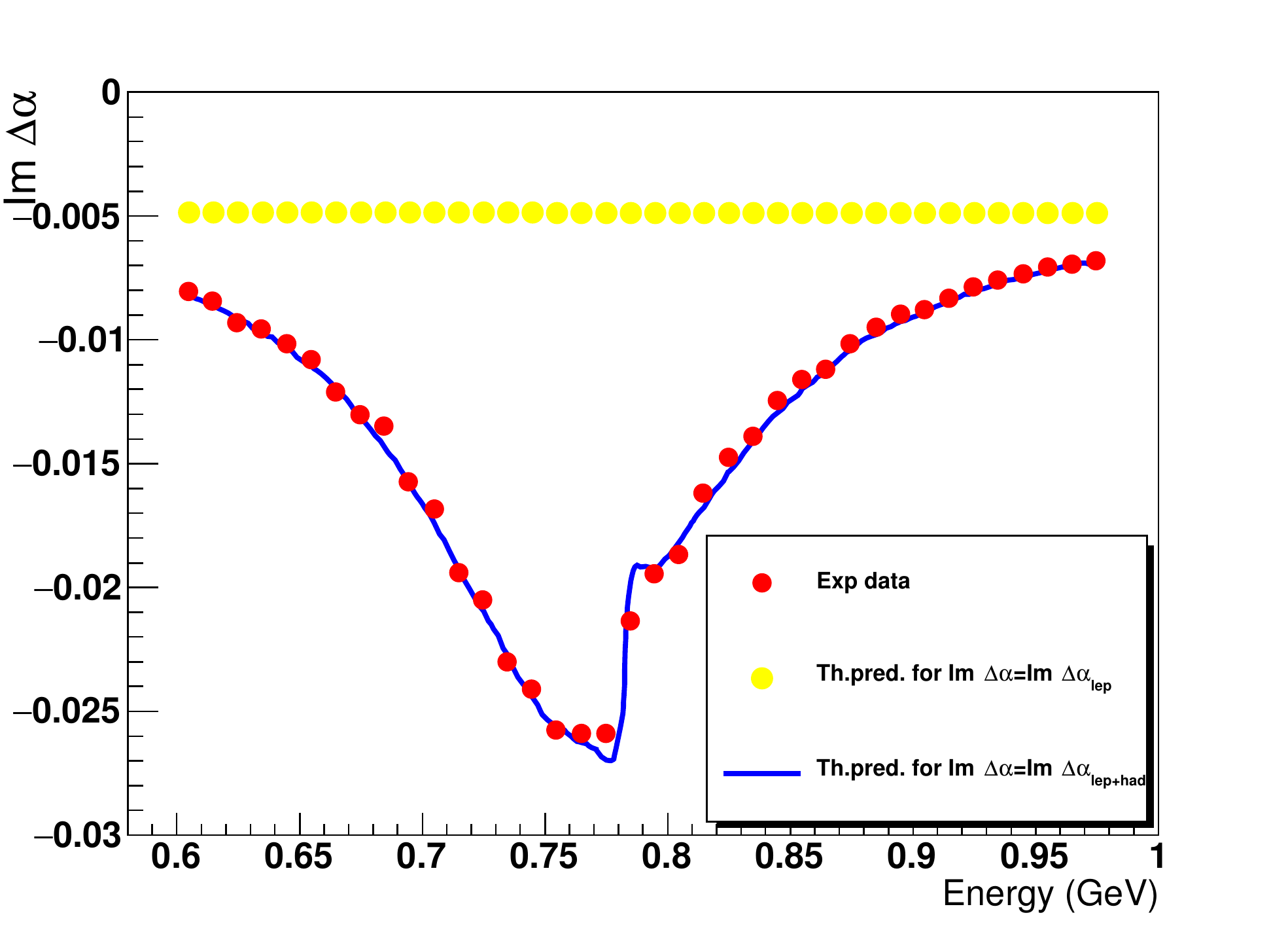}
\caption{$\rm Im\,\Delta \alpha$ extracted from the KLOE data 
compared with the values provided by {\tt alphaQED} routine (without the KLOE data) for
 $\rm Im\,\Delta \alpha=\rm Im\,\Delta \alpha_{lep}$ (yellow points) and $\rm Im\,\Delta \alpha=\rm Im\,\Delta \alpha_{lep+had}$ only for $\pi\pi$ channels (blue solid line).
 }
\label{im_delta_alpha}
\end{center}
\end{figure}

The results obtained for the $2\pi$ contribution to the imaginary part of $\Delta\alpha(s)$ by using the KLOE pion form factor measurement\cite{Venanzo},
are shown in Fig.~\ref{im_delta_alpha} and
compared 
%with the theoretical prediction made by F. Jergerlehner
with the values given by the $R_{had}(s)$ compilation of Ref.~\cite{fj} using only the $2\pi$ channel, with the KLOE data removed (to avoid correlations).  
Table~\ref{tab_alpha} gives the $2\pi$ contribution to $\rm Im\,\Delta\alpha(s)$ with statistical and systematic errors.

\begin{figure}[htp!]
\begin{center}
\includegraphics[width=10.8cm]{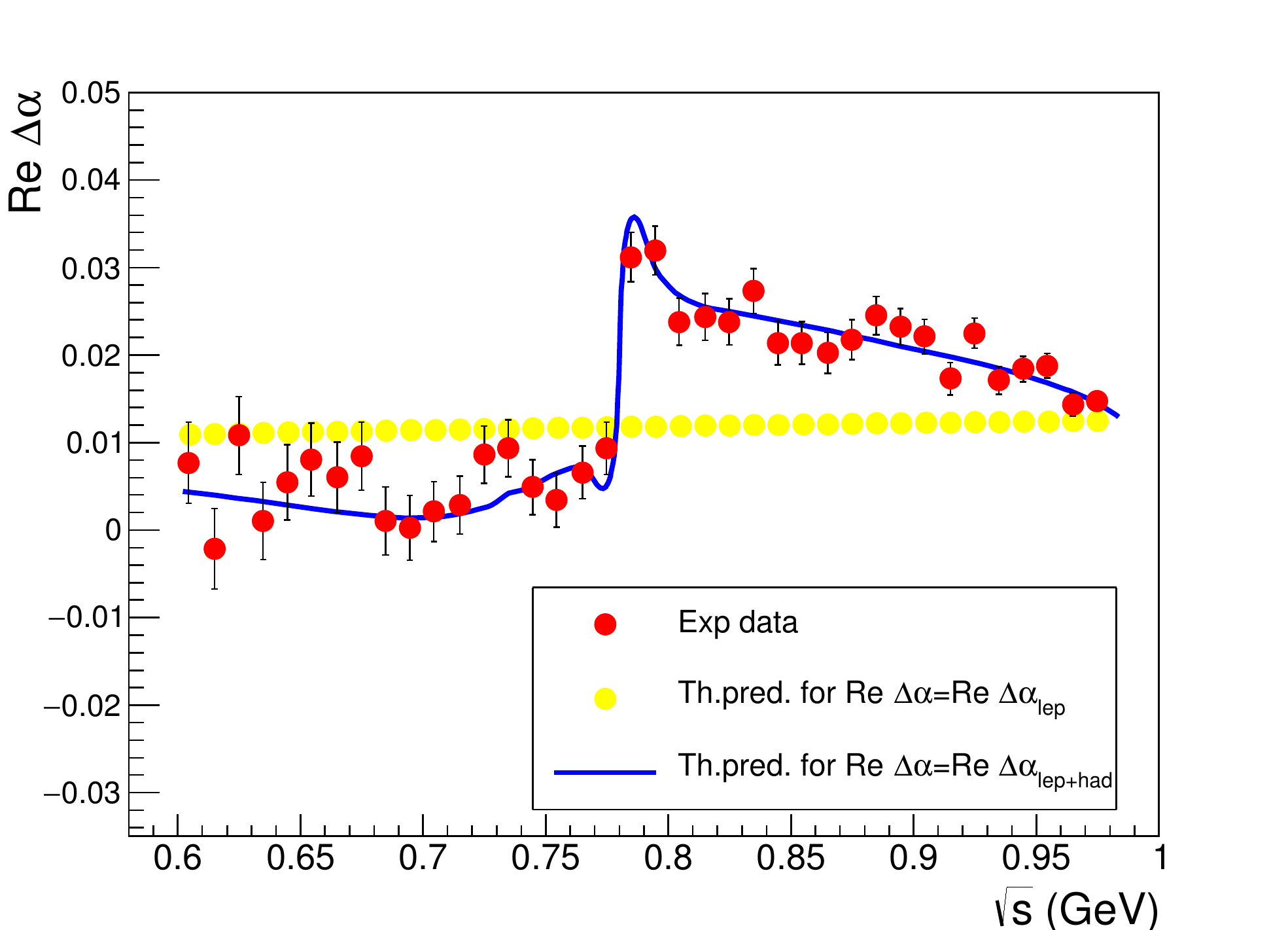}
\caption{$\rm Re\,\Delta \alpha$ extracted from the experimental data with only the statistical error included compared with the {\tt alphaQED} prediction (without the KLOE data) 
 when $\rm Re\,\Delta \alpha=\rm Re\,\Delta \alpha_{lep}$ (yellow points) and $\rm Re\,\Delta \alpha=\rm Re\,\Delta \alpha_{lep+had}$ (blue solid line). 
}
\label{re_delta_alpha_all}
\end{center}
\end{figure}

The extraction of the $\rm Re\,\Delta \alpha$ has been performed using the Eq. 
(\ref{re_delta_alpha})
and it is shown in Fig.~\ref{re_delta_alpha_all}. The experimental data with only the statistical error included have been compared 
%with the theoretical prediction performed by F. Jegerlehner  
with the {\tt alphaQED} prediction when $\rm Re\,\Delta \alpha=\rm Re\,\Delta \alpha_{lep}$ (yellow points in the colour Figure) and $\rm Re\,\Delta \alpha=\rm Re\,\Delta \alpha_{lep+had}$ (dots with solid line). 
The Re $\Delta\alpha(s)$ values with statistical and systematic errors are given in Table~\ref{tab_alpha}. The systematic errors include
%the contributions from missing hadronic channels which were not included in the evalution of Im $\Delta\alpha(s)$ 
the missing hadronic contributions (3$\pi$, 4$\pi$,...) which were not included in the evaluation of $\rm Im\,\Delta\alpha(s)$.
%gives the $2\pi$ contribution to Im $\Delta\alpha(s)$ with statistical and systematic errors.
%The difference at the $\omega$ peak is mostly given by the missing 3$\pi$ contribution.\\
As can be seen, an excellent agreement for $\rm Re\,\Delta\alpha(s)$ has been obtained with the data-based compilation.
% both Real and Imaginary part of $\Delta\alpha$ .

\begin{table}[hp!]
\begin{center}\tiny
%\begin{tabular}{|c|c|c|} \hline
\begin{tabular}{|c|c|c|c|} \hline
$\sqrt{s}$(GeV) & $\vert \frac{\alpha(s)}{\alpha(0)}\vert^2$& Re $\Delta\alpha$ $\cdot$ 10$^{-3}$ & 
Im $\Delta\alpha$ $\cdot$ 10$^{-3}$ \\
\hline
0.605  &   1.015 $\pm$ 0.010$\pm$0.011   & 7.7$\pm$4.7$\pm$5.4 & -8.09$\pm$0.08$\pm$0.03\\
0.615  &   0.996 $\pm$ 0.009$\pm$0.014   & -2.2$\pm$4.6$\pm$6.9 & -8.47$\pm$0.09$\pm$0.03\\
0.625  &   1.022 $\pm$ 0.009$\pm$0.014   & 10.7$\pm$4.5$\pm$6.8 & -9.35$\pm$0.10$\pm$0.04\\
0.635  &   1.002 $\pm$ 0.009$\pm$0.012   & 1.0$\pm$4.4$\pm$6.1 & -9.61$\pm$0.10$\pm$0.04\\
0.645  &   1.011 $\pm$ 0.009$\pm$0.012   & 5.4$\pm$4.3$\pm$6.0 & -10.20$\pm$0.12$\pm$0.04\\
0.655  &   1.016 $\pm$ 0.009$\pm$0.013   & 7.9$\pm$4.2$\pm$6.1 & -10.84$\pm$0.13$\pm$0.04\\
0.665  &   1.012 $\pm$ 0.008$\pm$0.011   & 6.0$\pm$4.1$\pm$5.6 & -12.15$\pm$0.15$\pm$0.06\\
0.675  &   1.017 $\pm$ 0.008$\pm$0.012   & 8.3$\pm$4.0$\pm$5.9 & -13.04$\pm$0.17$\pm$0.07\\
0.685  &   1.002 $\pm$ 0.008$\pm$0.011   & 0.9$\pm$3.9$\pm$5.4 & -13.51$\pm$0.17$\pm$0.04\\
0.695  &   1.000 $\pm$ 0.007$\pm$0.010   & 0.2$\pm$3.7$\pm$4.9 & -15.75$\pm$0.21$\pm$0.07\\
0.705  &   1.004 $\pm$ 0.007$\pm$0.010   & 2.1$\pm$3.4$\pm$5.1 & -16.89$\pm$0.23$\pm$0.07\\
0.715  &   1.005 $\pm$ 0.007$\pm$0.010   & 2.7$\pm$3.3$\pm$5 & -19.46$\pm$0.26$\pm$0.09\\
0.725  &   1.017 $\pm$ 0.007$\pm$0.010   & 8.6$\pm$3.3$\pm$4.9 & -20.54$\pm$0.28$\pm$0.11\\
0.735  &   1.018 $\pm$ 0.007$\pm$0.010   & 9.3$\pm$3.3$\pm$5.1 & -23.04$\pm$0.33$\pm$0.11\\
0.745  &   1.009 $\pm$ 0.006$\pm$0.010   & 4.8$\pm$3.2$\pm$4.7 & -24.15$\pm$0.34$\pm$0.23\\
0.755  &   1.006 $\pm$ 0.006$\pm$0.010   & 3.3$\pm$3.1$\pm$5.1 & -25.76$\pm$0.37$\pm$0.25\\
0.765  &   1.013 $\pm$ 0.006$\pm$0.010   & 6.5$\pm$3.1$\pm$5.1 & -25.89$\pm$0.37$\pm$0.25\\
0.775  &   1.018 $\pm$ 0.006$\pm$0.010   & 9.2$\pm$3.0$\pm$4.8 & -25.93$\pm$0.36$\pm$0.51\\
0.785  &   1.065 $\pm$ 0.007$\pm$0.011   & 31.1$\pm$3.0$\pm$4.9 & -21.36$\pm$0.27$\pm$0.69\\
0.795  &   1.066 $\pm$ 0.007$\pm$0.011   & 31.8$\pm$3.0$\pm$5.0 & -19.49$\pm$0.22$\pm$0.16\\
0.805  &   1.049 $\pm$ 0.006$\pm$0.011   & 23.8$\pm$2.8$\pm$5.0 & -18.69$\pm$0.22$\pm$0.17\\
0.815  &   1.050 $\pm$ 0.006$\pm$0.011   & 24.2$\pm$2.8$\pm$5.1 & -16.2$\pm$0.17$\pm$0.12\\
0.825  &   1.049 $\pm$ 0.006$\pm$0.011   & 23.6$\pm$2.7$\pm$5.0 & -14.79$\pm$0.15$\pm$0.11\\
0.835  &   1.057 $\pm$ 0.006$\pm$0.011   & 27.2$\pm$2.7$\pm$5.0 & -13.93$\pm$0.13$\pm$0.1\\
0.845  &   1.044 $\pm$ 0.006$\pm$0.011   & 21.3$\pm$2.6$\pm$5.0 & -12.49$\pm$0.11$\pm$0.08\\
0.855  &   1.044 $\pm$ 0.005$\pm$0.011   & 21.4$\pm$2.5$\pm$4.9 & -11.65$\pm$0.09$\pm$0.06\\
0.865  &   1.041 $\pm$ 0.005$\pm$0.011   & 20.2$\pm$2.5$\pm$5.0 & -11.25$\pm$0.09$\pm$0.05\\
0.875  &   1.044 $\pm$ 0.005$\pm$0.011   & 21.6$\pm$2.4$\pm$5.1 & -10.16$\pm$0.07$\pm$0.04\\
0.885  &   1.050 $\pm$ 0.005$\pm$0.011   & 24.4$\pm$2.3$\pm$4.9 & -9.53$\pm$0.06$\pm$0.03\\
0.895  &   1.048 $\pm$ 0.005$\pm$0.011   & 23.1$\pm$2.2$\pm$5 & -9.03$\pm$0.05$\pm$0.03\\
0.905  &   1.045 $\pm$ 0.004$\pm$0.011   & 22.0$\pm$2.1$\pm$5.1 & -8.81$\pm$0.05$\pm$0.02\\
0.915  &   1.035 $\pm$ 0.004$\pm$0.011   & 17.2$\pm$1.9$\pm$5.3 & -8.35$\pm$0.04$\pm$0.02\\
0.925  &   1.046 $\pm$ 0.004$\pm$0.011   & 22.3$\pm$1.8$\pm$5.1 & -7.89$\pm$0.03$\pm$0.02\\
0.935  &   1.035 $\pm$ 0.003$\pm$0.011   & 17.0$\pm$1.7$\pm$5.1 & -7.62$\pm$0.03$\pm$0.01\\
0.945  &   1.038 $\pm$ 0.003$\pm$0.010   & 18.3$\pm$1.5$\pm$4.8 & -7.33$\pm$0.02$\pm$0.01\\
0.955  &   1.039 $\pm$ 0.003$\pm$0.010   & 18.8$\pm$1.4$\pm$4.8 & -7.13$\pm$0.02$\pm$0.01\\
0.965  &   1.029 $\pm$ 0.003$\pm$0.010   & 14.2$\pm$1.3$\pm$4.8 & -6.94$\pm$0.02$\pm$0.01\\
0.975  &   1.030 $\pm$ 0.002$\pm$0.010   & 14.6$\pm$1.1$\pm$4.7 & -6.82$\pm$0.02$\pm$0.01\\

\hline
\end{tabular}
\end{center}
\caption{$\bigl\lvert \frac{\alpha(s)}{\alpha(0)}\bigr\rvert^2, \rm Re\,\Delta\alpha, \rm Im\,\Delta\alpha$ values in 0.01 GeV intervals are reported; the first error is statistical, the second is systematic. 
The hadronic contribution to $\rm Im\,\Delta\alpha$ includes only the 2$\pi$ channel from Ref.~\cite{Venanzo}. 
}
\label{tab_alpha}
\end{table}

\clearpage

\section{Fit of $\rm Re\,\Delta \alpha$ and extraction of 
$BR(\omega\to\mu^+\mu^-)BR(\omega\to e^+e^-)$}

We fit $\rm Re\,\Delta \alpha$ by a sum of the leptonic and hadronic contributions, where
the hadronic contribution is parametrized as a sum of the $\rho(770)$, $\omega(782)$ and $\phi(1020)$ resonance components and a non-resonant term. 
We use a Breit-Wigner description for the $\omega$ and $\phi$ resonances~\cite{Iwao:1974ib,actis,fred}:
\begin{equation}
\rm Re\,\Delta \alpha_{V=\omega,\phi} =
\frac{3\sqrt{BR(V\to e^+e^-)\cdot BR(V\to\mu^+\mu^-)}}{\alpha M_V}\frac{s(s-M^2_V)\Gamma_V}{(s-M^2_V)^2+s\Gamma^2_V},
\end{equation}
where $M_V$ and $\Gamma_V$ are the mass and the total width of the mesons $V=\omega$ and $\phi$.
For the $\rho$ we use a Gounaris-Sakurai parametrization $BW^{GS}_{\rho(s)}$~\cite{Gounaris:1968mw,Akhmetshin:2001ig} of the pion form factor, where we neglect the interference with the $\omega$ and the higher excited states of the $\rho$:

\begin{equation}
F_{\pi}(s)= BW^{GS}_{\rho(s)} = \frac{M^2_\rho(1+d \Gamma_\rho/M_{\rho})}{M^2_{\rho}-s+f(s)-iM_{\rho}\Gamma_{\rho}(s)}
\end{equation}

The terms $d$ and $f(s)$ are described in Ref.~\cite{Akhmetshin:2001ig}.
As it will be shown in the following, this approximation turns out to be appropriate given the limited statistics of the data. 
In particular, the inclusion of the energy dependence on the total widths of $\omega$ and $\phi$ resonances~\cite{Achasov:2000am} gives negligible contributions.
The non-resonant term has been parametrized as a first-order polynomial $p_0+p_1\sqrt s$.\\
The following parameters have been fixed to the PDG values~\cite{pdg}: $\Gamma_{\omega}=(8.49\pm0.08)$ MeV, $M_{\phi}=(1019.461\pm0.019)$ MeV, 
$\Gamma_{\phi}=(4.266\pm0.031)$ MeV, and $BR(\phi\to\e^+e^-)BR(\phi\to\mu^+\mu^-)=(8.5^{+0.5}_{-0.6})\cdot10^{-8}$. \\
\begin{figure} [h]
\centerline{
\includegraphics[width=28pc]{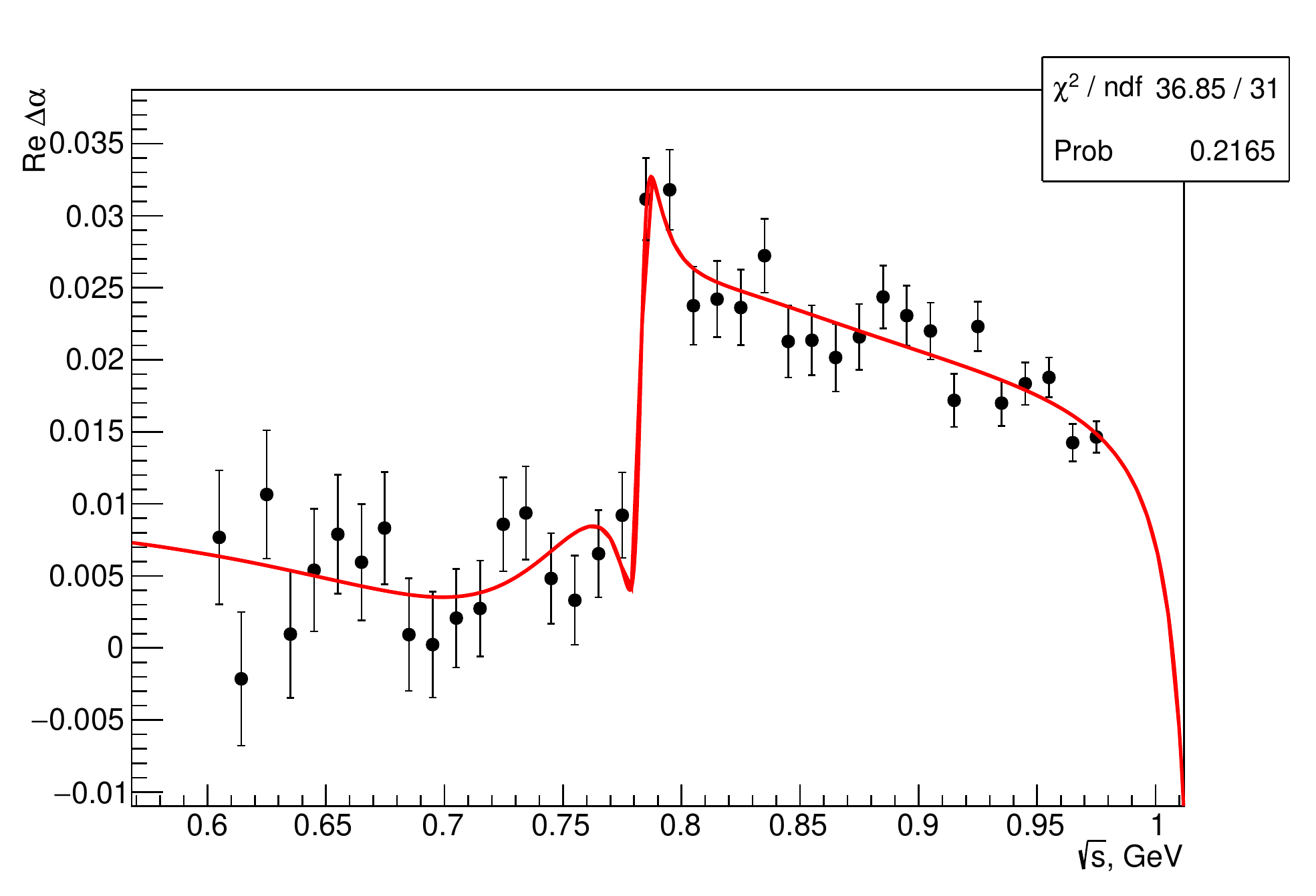}}
\caption{Fit of the $\rm Re\,\Delta \alpha$ data. Only statistical errors are shown.}
\label{brfit_plb}
\end{figure}

\noindent Results of the fit are shown in Fig.~\ref{brfit_plb} and compared in Table~\ref{tab:fit} (second column) with the corresponding values from PDG~\cite{pdg}. Only statistical errors are reported. \\
\begin{table}
\begin{center}
\scalebox{0.7}{
\begin{tabular}{||l|c|c|c||}
\hline\hline
Parameter & Result from the fit &  Result from the fit with $\rho-\omega$ interf.& PDG\\
\hline\hline
$M_{\rho}$, MeV & $775\pm6$ &  $778\pm7$ & $775.26\pm0.25$ \\
$\Gamma_{\rho}$, MeV & $146\pm9$ & $147\pm 10$ & $147\pm 0.9$\\
$M_{\omega}§$, MeV & $782.7\pm1.1$ & $783.4\pm0.8$ &$782.65\pm0.12$ \\
$BR(\omega\to\mu^+\mu^-)BR(\omega\to e^+e^-)$ & $(4.3\pm 1.8)\cdot10^{-9}$  
&  $(4.4\pm 1.8)\cdot10^{-9}$ & $(6.5\pm 2.3)\cdot10^{-9}$\\
$\chi^2/ndf$ & 1.19 & 1.15 & - \\
\hline\hline
\end{tabular}
}
\vspace{0.2cm}
\caption{Results from the fit of $\rm Re\,\Delta \alpha$ compared with the world average values (PDG~\cite{pdg}). Second (third) column: without (with) the $\rho-\omega$ interference. Only statistical errors are reported for the fit values.}
\label{tab:fit}
\end{center}
\end{table}
\noindent The parameters of the non-resonant term are consistent with zero within the statistical uncertainties: $p_0=(2.4\pm4.5)\cdot 10^{-3}$, $p_1=(-2.8\pm5.3)\cdot 10^{-3}$. The $\chi^2/ndf$ of the fit is $36.85/31=1.19$.

To study %the effect %of the interference between $\rho$ and $\omega$ $\rm Re\,\Delta \alpha$ 
the effect of the $\rho-\omega$ interference in estimating $\Delta \alpha$, 
an additional term 
$\delta\frac{s}{M^2_{\omega}}BW_{\omega}(s)BW^{GS}_{\rho}$ has been included in the fit. 
Results are shown in the third column of Table~\ref{tab:fit} where we fix $|\delta|= 1.45\cdot10^{-3}$ and  $arg$ $\delta= 10.2^{\deg}$~\cite{fitKLOE}.
As it can be seen, results with the interference term 
are well within the statistical uncertainties, and in the following we will use the results without the interference term.\\
By including the systematic errors (%taking into account the effect and the correlation 
taking also into account the correlations
of the systematic uncertainties on the parameters of the fit,
and the uncertainty of the PDG values for fixed parameters)
%by allowing the data to move for the systematic uncertainties and including also the uncertainty on the PDG values for fixed parameters) 
the product of the branching fractions reads:
\begin{equation}
BR(\omega\to\mu^+\mu^-)BR(\omega\to e^+e^-) = (4.3\pm1.8\pm2.2)\cdot 10^{-9},
\end{equation}
where the first error is statistical and the second systematic. 
By multiplying by the phase space factor % $\xi$
%\begin{equation}
$\xi =\Big(1+2\frac{m^2_\mu}{m^2_{\omega}}\Big)\Big(1-4\frac{m^2_\mu}{m^2_{\omega}}\Big)^{1/2}$
%\end{equation}
 and assuming lepton universality, $BR(\omega\to\mu^+\mu^-)$ can be extracted:
\begin{equation}
BR(\omega\to \mu^+\mu^-)=(6.6\pm1.4_{stat}\pm1.7_{syst})\cdot 10^{-5} 
\end{equation}
compared to $BR(\omega\to\mu^+\mu^-) =(9.0\pm3.1)\cdot 10^{-5}$ from PDG~\cite{pdg}.\\

\section{Conclusions}
\label{conclusions}
We have measured the hadronic contribution to the running of the effective QED coupling constant $\alpha(s)$ using the differential cross section $d\sigma(e^+e^- \rightarrow \mu^+\mu^- \gamma)/d \sqrt{s}$ in the region $0.6<\sqrt s< 0.975$ GeV, with the photon emitted in the initial state.
Our results  show a clear contribution of the $\rho-\omega$ resonances to the photon propagator, which results in a  
more than 5$\sigma$ significance of the hadronic contribution to the running of $\alpha(s)$. This is the strongest direct evidence achieved in both time- and space-like regions by a single experiment.
For the first time the real and imaginary parts of $\Delta\alpha(s)$ have also been extracted.
From a fit of the real part of $\Delta\alpha(s)$ and assumming the lepton universality 
the branching ratio $BR(\omega\to\mu^+\mu^-) = (6.6\pm1.4_{stat}\pm1.7_{syst})\cdot 10^{-5} $ has been obtained.

\section{Acknowledgments}
We thank F.~Ignatov and C.~M.~Carloni~Calame for useful discussions.
%We thank F.~Ignatov for many useful discussion and C.M.Carloni Calame for clarifications on the fit section.\\
We warmly thank our former KLOE colleagues for the access to the data collected during the KLOE data taking campaign.
We thank the DA$\Phi$NE team for their efforts in maintaining low background running conditions and their collaboration during all data taking. We want to thank our technical staff: 
G.F. Fortugno and F. Sborzacchi for their dedication in ensuring efficient operation of the KLOE computing facilities; 
M. Anelli for his continuous attention to the gas system and detector safety; 
A. Balla, M. Gatta, G. Corradi and G. Papalino for electronics maintenance; 
M. Santoni, G. Paoluzzi and R. Rosellini for general detector support; 
C. Piscitelli for his help during major maintenance periods. 
This work was supported in part by the EU Integrated Infrastructure Initiative Hadron Physics Project under contract number RII3-CT- 2004-506078; by the European Commission under the 7th Framework Programme through the `Research Infrastructures' action of the `Capacities' Programme, Call: FP7-INFRASTRUCTURES-2008-1, Grant Agreement No. 227431; by the Polish National Science Centre through the Grants No.\
%0469/B/H03/2009/37, 
%0309/B/H03/2011/40, 
%DEC-2011/03/N/ST2/02641,   % J.Zdebik - out 4 Jan 2015
%2011/01/D/ST2/00748,
2011/03/N/ST2/02652,
2013/08/M/ST2/00323,
2013/11/B/ST2/04245,
2014/14/E/ST2/00262,
2014/12/S/ST2/00459.
%0469/B/H03/2009/37, 
%0309/B/H03/2011/40, 
%DEC-2011/03/N/ST2/02641,   % J.Zdebik - out 4 Jan 2015
%2011/01/D/ST2/00748,

%\end{linenumbers}
\end{document}